\documentstyle[11pt,newpasp,twoside,epsf]{article}
\def\edcomment#1{\iffalse\marginpar{\raggedright\sl#1\/}\else\relax\fi}
\reversemarginpar

\begin{document}

 \title{Astrometric demands of fibre-input spectrographs for wide-field
quasar redshift surveys}
 \author{Peter R.\ Newman}
 \affil{Centre for Astrophysics, Univ. of Central Lancashire, Preston PR1
2HE}

 \begin{abstract}
 A potential problem with using wide-field multi-object spectrographs for
quasar redshift surveys is the introduction of position-dependent
selection effects into the survey catalogue because of variations in the
observed signal-to-noise ratio (SNR) across the survey area.  A function
relating signal losses in instruments using optical fibre input to the
fibre-to-image position error, the fibre diameter and the seeing
conditions is given and plotted for a range of typical values of fibre
sizes and seeing found in current instruments. 
 \end{abstract}

 \section{Why this is interesting}

 The new era of wide-field redshift surveys, exemplified by 2df and SDSS
(see Boyle et al.\ and Lupton et al., these proceedings), relies on
multi-object spectrographs to observe thousands of candidates per night. 
Constructing homogeneous catalogues from such data requires automated
analysis to achieve consistent and reliable spectral classifications and
redshift measurements.  The probability of correct results depends on the
SNR achieved during observations, which is strongly affected by how much
of each target image actually falls on a spectrograph aperture.  Compared
to single-object spectrographs, multi-object instruments thus introduce
problems in reaching a consistent minimum SNR for faint objects across the
survey.  For fibre-input instruments observing point sources, the seeing,
fibre diameter and fibre-to-image position error are the important
parameters. Typical galaxy images are $\ga 4$~$h$~arcsec across, giving
some tolerance to fibre positioning errors, but quasar images are usually
seeing-limited at $\sim0.5$--$2.0$~arcsec, demanding greater precision. 
Fibre-to-image offsets will clearly reduce the recorded signal and may
also vary systematically across the field, leading to position-dependent
selection.

 \section{Signal reduction due to fibre position errors}

Most fibre spectrographs use blind positioning of individual fibres in (or
close to) the telescope focal surface before observation, and employ a
small number of coherent fibre bundles positioned on guide star images to
centre the telescope on the desired field, adjust fibre mount plate
rotation and track the field during exposure.  Exposures are typically
$45$--$90$ min.  Sources of fibre-to-image position errors in such systems
include the fibre positioning robot or plate drilling machine, variations
in differential atmospheric refraction and dispersion during exposure,
image scale, focal surface and field distortion mapping, temperature
changes in the fibre mount plate, and plate rotation and guiding errors.
For typical 2\deg-field instruments, the combination of these errors can
give an r.m.s.\ fibre position uncertainty of $\sim0.5$~arcsec (see e.g.\
Bridges 2000, Cuby 1994, Shectman 1993, Barden \& Massey 1988).  This is
in addition to any input catalogue astrometric uncertainty which may be of
similar size, particularly for surveys based on input catalogues
constructed from Schmidt plates.

Quasars images are well approximated by a 2-D Gaussian point-spread
function (PSF), with a size determined by the seeing disk (Brodie et al. 
1988).  For a fibre of radius $R$ displaced by $x_0$ from the centre of a
symmetric PSF of variance $\sigma^2$, the fraction of the image energy
entering the fibre may be calculated from the integral of the PSF over a
circular aperture, giving
 \begin{equation}
 f(x_0) = \frac{1}{\sigma\sqrt{2\pi}}\int_{-R}^{R}
          \exp\left(\frac{-(x-x_0)^2}{2\sigma^2}\right) 
          \mbox{erf}\left(\frac{\sqrt{R^2-x^2}}{\sqrt{2}
          \,\sigma}\right)\,dx
 \label{eq:aper_energy_simple}
 \end{equation}
 where $\mbox{erf}$ is the error function (Newman 1999).  With measures of
the seeing during observations, this function may be of use in determining
the effective limiting magnitude of each of the individual exposures that
make up a survey. 

 \section{Conclusions and further work}

Fig.\ 1 shows how $f(x_0)$ varies with $x_0$ for a range of typical fibre
diameters and seeing conditions, and demonstrates how instruments with
small fibre diameters place tremendous demands on fibre positioning
precision and input catalogue astrometry.  Fibres with diameters
comparable to the seeing disk FWHM greatly reduce the signal even from
well-centred images. Surveys constructed from tiled observations made
under varying conditions may thus suffer from a very complicated
position-dependent faint-end selection effect.  The ultimate aim of this
work is to characterize the position dependence of this selection and
hence the catalogue incompleteness in such surveys, as well as quantifying
a factor that should be considered in the design of future fibre-input
spectrographs and quasar surveys made using them.

\acknowledgements Part of this work was supported by PPARC grant
B95300682.  Thanks to Terry Bridges for several useful discussions.

 \begin{figure}
 \plotone{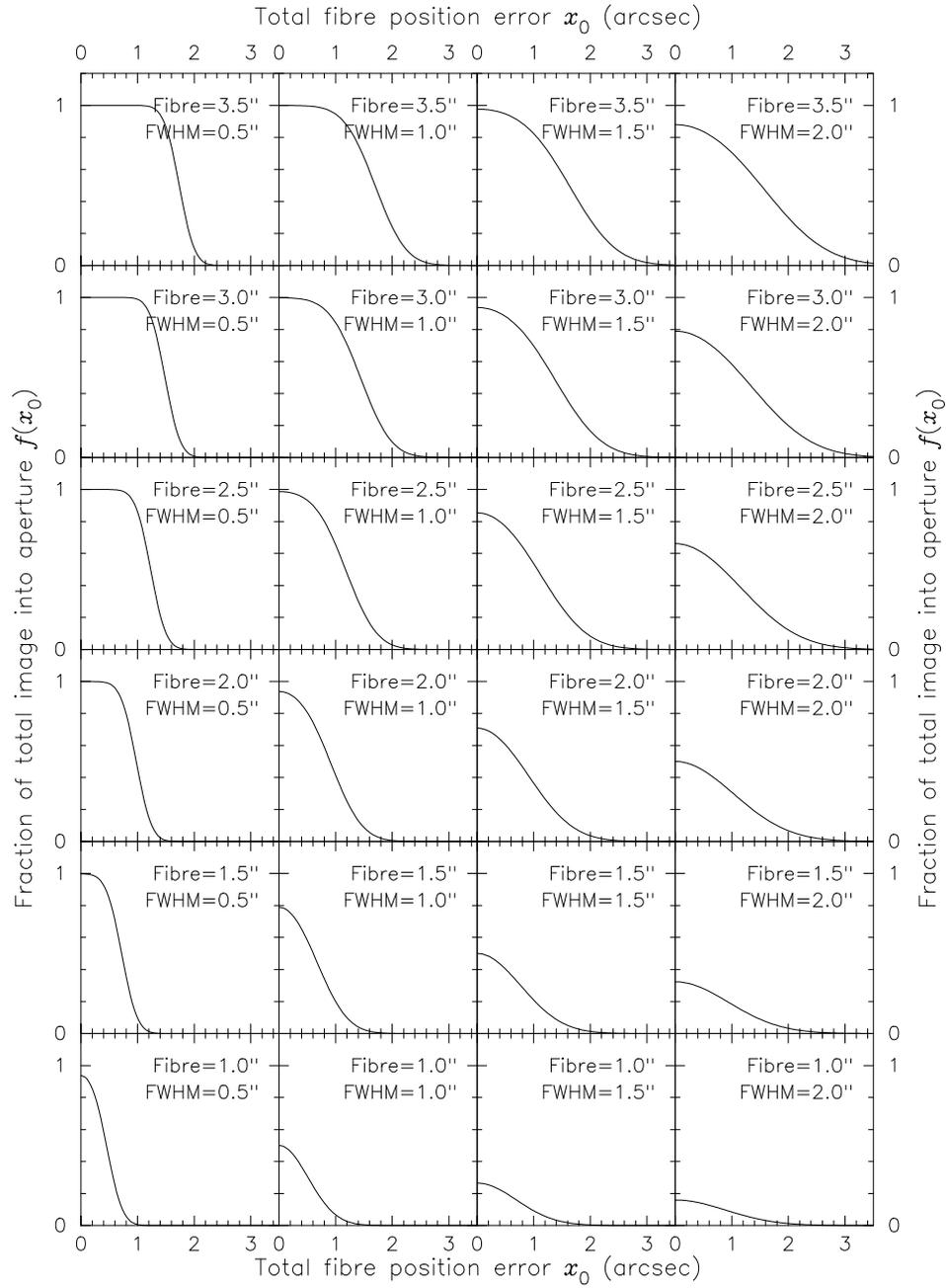}
 \caption{
 $f(x_0)$ for a range of fibre diameters and seeing FWHM.
 }
 \end{figure}

\end{document}